\documentclass[slac_one]{revtex4}
\pdfoutput=1
\usepackage{graphicx}
\usepackage{fancyhdr}
\usepackage{subfigure}
\def\cms{cm$^{-2}$~s$^{-1}$}
\def\TeV{\ifmmode {\mathrm{\ Te\kern -0.1em V}}\else
                   \textrm{Te\kern -0.1em V}\fi}%

\def\GeV{\ifmmode {\mathrm{\ Ge\kern -0.1em V}}\else
                   \textrm{Ge\kern -0.1em V}\fi}%

\def\MeV{\ifmmode {\mathrm{\ Me\kern -0.1em V}}\else
                   \textrm{Me\kern -0.1em V}\fi}%

\def\ipb{\mbox{pb$^{-1} $} }
\def\pT{\ensuremath{p_{\mathrm{T}}}} 
\def\kT{\ensuremath{k_{\mathrm{T}}}} 
\def\eT{\ensuremath{E_{\mathrm{T}}}} 
\def\s{\ensuremath{\sqrt{s } } } %

\begin{document}

\title{The ATLAS Trigger System Commissioning and Performance} 

%

\author{Andrew Hamilton on behalf of the ATLAS Collaboration}
\affiliation{Universit\'e de Gen\`eve, Section de Physique, 24 rue Ernest Ansermet, CH - 1211 Gen\`eve 4, Switzerland}

\begin{abstract}
The ATLAS trigger has been used very successfully to collect collision data during 2009 and 2010 LHC running at centre of mass energies of 900 GeV, 2.36 \TeV, and 7 \TeV.   This paper presents the ongoing work to commission the ATLAS trigger with proton collisions, including an overview of the performance of the trigger based on extensive online running.   We describe how the trigger has evolved with increasing LHC luminosity and give a brief overview of plans for forthcoming LHC running.\end{abstract}

\maketitle

\thispagestyle{fancy}

\section{The ATLAS Trigger}

The LHC~\cite{lhc} is a high-energy, high-intensity hadron collider built to study the Standard Model (SM) and search for physics beyond the SM.  ATLAS~\cite{atlas} is one of two general-purpose experiments located at the LHC.  At design luminosity ($10^{34}$~\cms) and collision energy (14 \TeV) the LHC will have a bunch-crossing rate of 40 MHz and deliver approximately $10^9$ proton-proton collisions per second to the ATLAS detector.   The average output bandwidth of the ATLAS data-acquisition system (DAQ) is about 200 Hz.   The rate reduction from 40MHz to 200Hz is achieved using the ATLAS trigger~\cite{trigger}, by selecting only the most interesting events for physics analysis.

The LHC commissioning with colliding proton beams began at \s = 900 \GeV~in November 2009 and continued at \s = 7 \TeV~in March 2010.  By the end of August 2010, ATLAS collected 3.2 \ipb of data with a peak luminosity of $1.05 \times 10^{31}$\cms.   The ATLAS experiment is using this first data to commission the experiment and produce the first performance~\cite{perf} and physics~\cite{minbias}\cite{jets} results. The ATLAS trigger system is designed to facilitate this initial phase of the experiment as well as meeting the more stringent requirements of increased luminosity and collision energy.

The ATLAS trigger is a three level trigger system.  The first level, \emph{Level 1} (L1), uses custom electronics to rapidly select collisions of interest.  The next two levels, \emph{Level 2} (L2) and \emph{Event Filter} (EF), are collectively known as the \emph{High Level Trigger} (HLT).  The HLT, a software based system running on farms of commercial computers, refines the selection of L1 with increased algorithm complexity and detector granularity at the cost of longer decision time. 

When a bunch-crossing is signaled by the LHC clock, reduced granularity detector data are sent to the L1 processing boards and the full granularity data are sent to front-end pipeline memories.  The L1 processing boards perform calorimeter clustering and energy summing, muon spectrometer track candidate identification, and minimum bias  trigger scintillator (MBTS) threshold discrimination.  The results from the L1 processing boards are combined in the central trigger processor, which accepts or rejects events based on \pT~and/or quantity of calorimeter or muon information.   The maximum latency of the L1 trigger is 2.5 $\mu$s, set by the length of the front-end pipeline.  The maximum output rate from L1 is 75 kHz; future upgrades could make this 100 kHz.   

In the case of an L1 accept, the ($\eta$, $\phi$) location of the calorimeter cluster or muon track, known as the \emph{Region-of-Interest} (RoI), is sent to L2.   The L2 uses fast, dedicated algorithms to accept or reject the event by performing reconstruction with full granularity data.  The amount of data to be transferred and processed in L2 is reduced to 2 - 4\% of the total data volume by using only the volume surrounding the RoI.  For the appropriate signatures, calorimeter or muon spectrometer information is combined with \emph{inner detector} (inside the solenoid) tracks.    The final size of the L2 processing farm will be about 500 multi-core processors; the average processing time will be about 40ms. The maximum L2 output rate is limited by the event building bandwidth to about 3kHz. 

After an L2 accept, the full event information is collected and sent to the EF. The EF has access to the full event data and the algorithms running are typically using the same software as the offline event reconstruction.  The final EF farm will be about 1800 multi-core processors; the average processing time will be about 4s. The maximum HLT output rate is limited to about 200Hz due to computing resource limitations for offline event processing. 

The complete trigger selection is defined by the trigger \emph{menu}, which is chosen when the DAQ is configured before each run.  A menu consists of many trigger \emph{chains} (typically several hundred), where each chain defines the L1 and HLT selection for a single physics signature, such as an electron with \pT $>$ 20~\GeV~or two muons with an invariant mass around 3.1~\GeV.

\section{Commissioning} 

The commissioning of the ATLAS trigger system began before LHC provided colliding beams by using Monte Carlo (MC) events and cosmic rays.   During the MC commissioning, simulated collisions were inserted into the readout system and processed by the HLT.   Triggering on cosmic rays was used to exercise the basic muon, inner detector tracking, and calorimeter clustering algorithms. The HLT was used to select a large sample of cosmic ray events to be used for inner detector tracking commissioning~\cite{idcomm}.  

For the LHC collision running in 2009 and 2010, the trigger was commissioned in several steps. During the initial period of  900 \GeV collisions, no HLT algorithms were running online.  L1 was extensively tested and HLT algorithms were exercised offline using the collision events within hours of being recorded.  Shortly ($\sim$weeks) thereafter, the HLT algorithms were validated and enabled online in \emph{transparent mode}.  In this mode, the HLT algorithms are run normally, but all events are accepted,  regardless of the HLT decision.  HLT objects used to make the trigger decision are recorded into the data stream for offline analysis to evaluate the HLT performance. 

For the first collisions at 7 \TeV~in March 2010, the HLT algorithms were initially disabled, but within two hours of data taking and fast offline processing, the HLT was turned on in transparent mode.  The low instantaneous luminosity in this period ($\sim 10^{27}$\cms) allowed ATLAS to run efficiently in transparent mode until end of May 2010 when the instantaneous luminosity was sufficient ($\sim 10^{29}$\cms) to require the HLT to be progressively activated.  The highest rate triggers were validated then put into \emph{active rejection mode} where they rejected events that did not satisfy the trigger chain's requirements.  In July 2010, when the instantaneous luminosity reached $\sim10^{30}$\cms, the physics menu was deployed.  The physics menu contains more triggers designed for physics analysis than the commissioning menu it replaced.  The deployment of the physics menu is one of the final stages of the ATLAS trigger commissioning. 

The trigger menu has evolved several times through the commissioning period. For the 900 \GeV~running period in 2009, the HLT had about 170 chains running in transparent mode.  For the first 7 \TeV~collisions this was increased to about 220 chains and evolved to a commissioning menu containing about 420 chains.  Many of the chains were only included to ensure all of the HLT algorithms are exercised even at low luminosity.  The first physics menu deployed contains about 390 chains.

\begin{figure}[!h]
\centering

\subfigure[ ~L1 jet trigger efficiency. ]{
	\includegraphics[width=8.0cm]{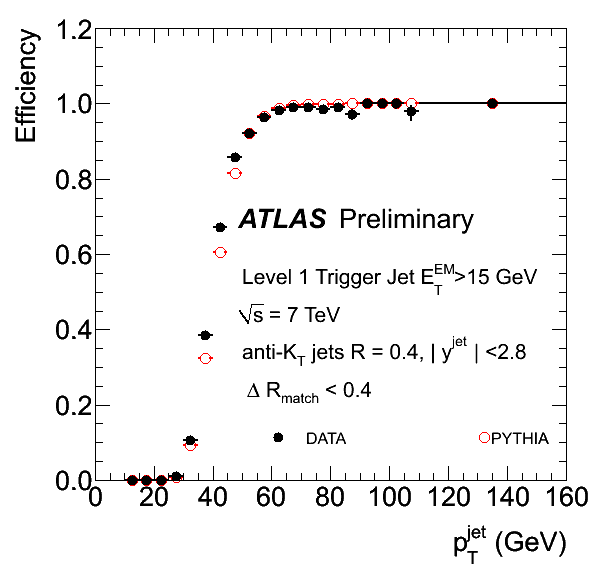}
	\label{fig:jet}
	}
\hspace{0.7cm}
\subfigure[ ~L1 trigger rates. ]{
	\includegraphics[width=8.0cm]{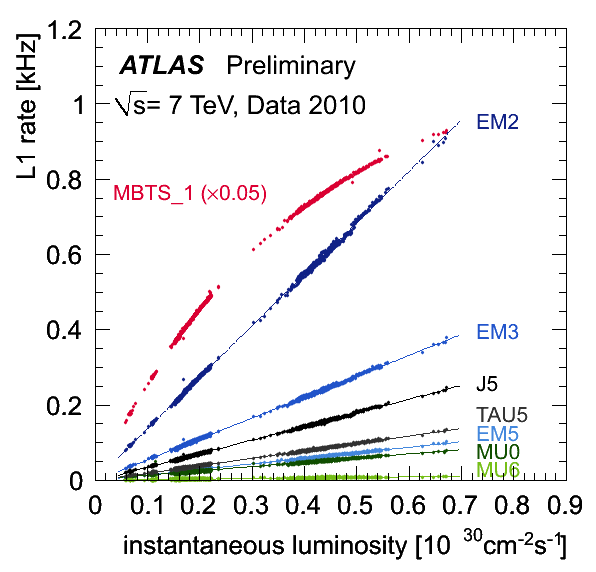}
	\label{fig:rates}
	}

\subfigure[ ~Beam spot determined by L2 tracking ]{
	\includegraphics[width=7.7cm]{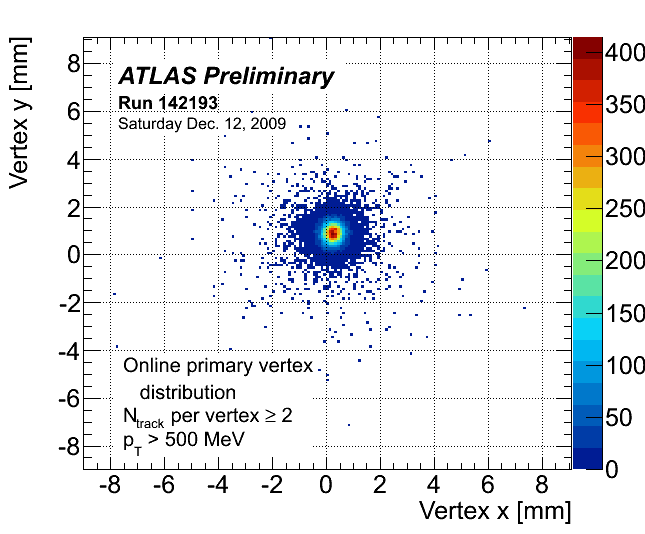}
	\label{fig:beamspot}
	}
\hspace{0.7cm}
\subfigure[ ~L1 muon efficiency in end-cap region. ]{
	\includegraphics[width=8.3cm]{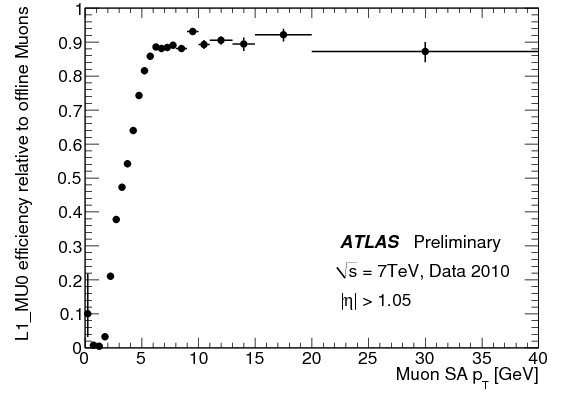}
	\label{fig:muon}
	}

\subfigure[ ~EF tracking efficiency.]{
	\includegraphics[width=8.3cm]{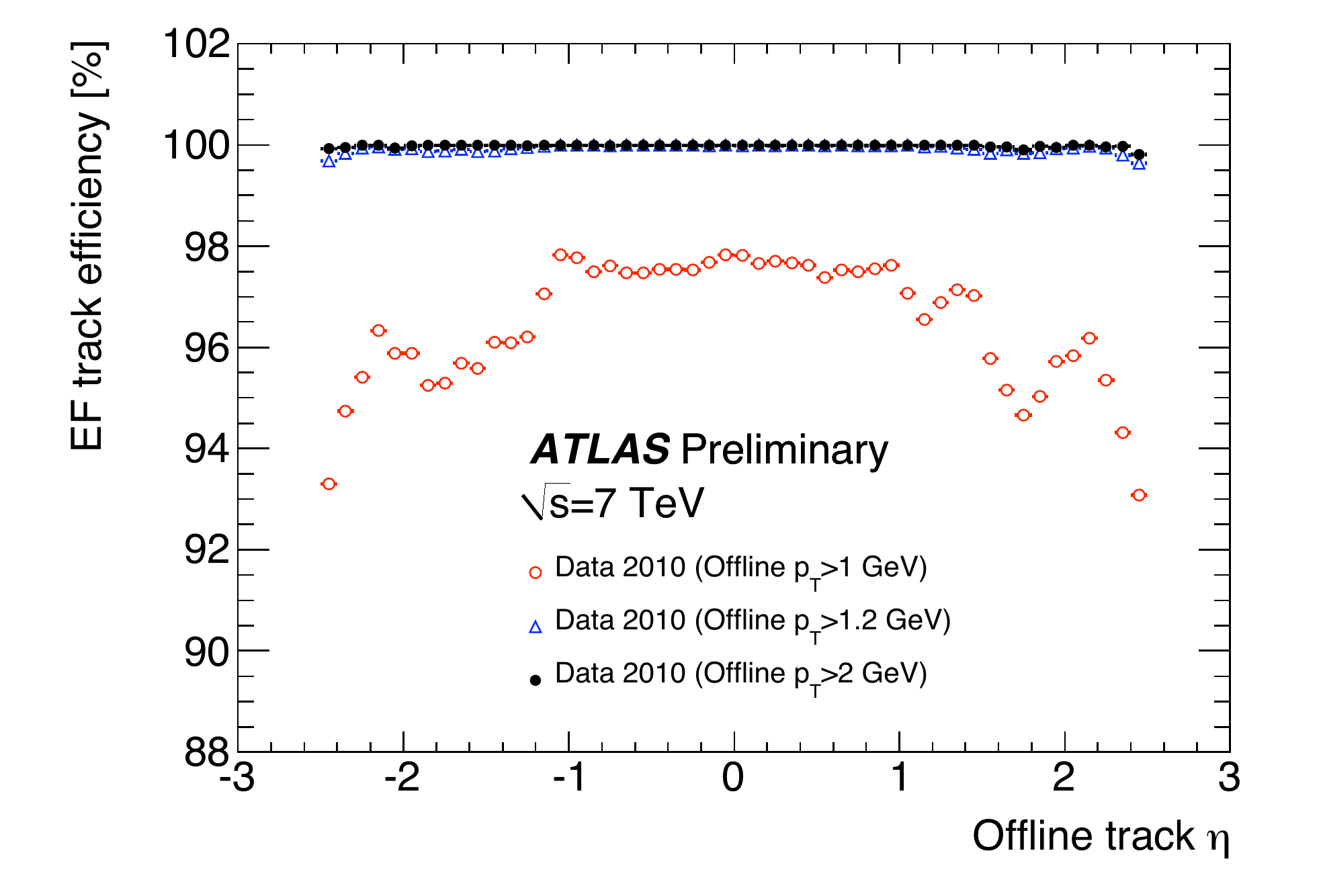}
	\label{fig:tracks}
	}
\hspace{0.7cm}
\subfigure[ ~L2 electromagnetic cluster \eT spectrum. ]{
	\includegraphics[width=7.7cm]{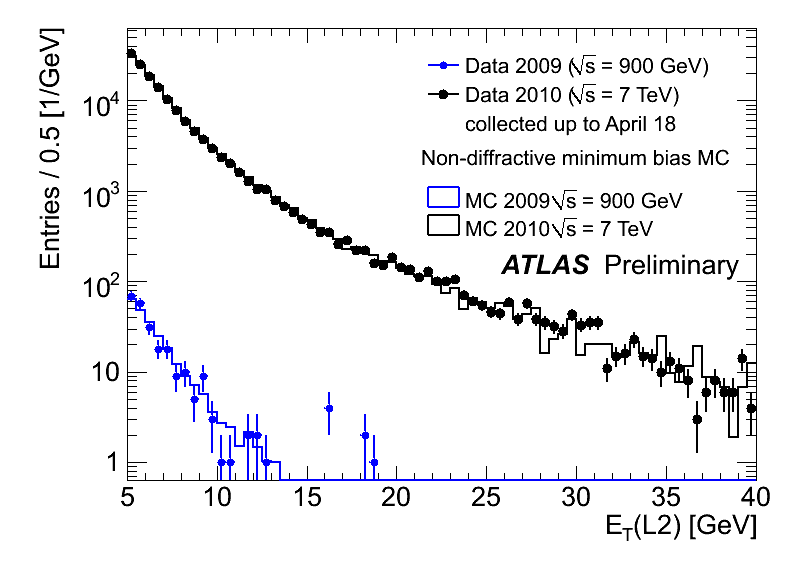}
	\label{fig:em}
	}
\caption{Examples of trigger performance plots.}
 \label{fig:plots}
\end{figure}

\section{Performance}

The trigger performance of all the major ATLAS subsystems has been measured and results are as expected from simulation.  Figure~\ref{fig:plots} provides examples of these performance checks for various detector systems.  For more detailed discussion of the trigger performance with respect to reconstructed physics objects, please refer to some of the other papers in these proceedings~\cite{mb}-\cite{muon}.

Figure~\ref{fig:jet} shows the 15 GeV L1 jet trigger efficiency for jets identified offline using the anti-\kT~(R=0.4) algorithm as a function of the calibrated offline jet \pT.  The efficiency was extracted using a data sample triggered by the MBTS and the energy of the jets was evaluated at electromagnetic scale of the calorimeter. 

Figure~\ref{fig:rates} shows the unprescaled L1 trigger rates growing as expected with instantaneous luminosity;  electromagnetic triggers (\eT~thresholds of 2, 3, and 5 \GeV), muon triggers (\pT~thresholds of 0 and 6 \GeV), a tau trigger (\eT~threshold of 5 \GeV), a jet trigger (\eT~threshold of 5 \GeV) and a trigger requiring a single hit in one of the minimum bias trigger scintillators (MBTS\_1).  In the figure, the MBTS\_1 rate is scaled down by a factor of 20.  Each dot represents a measurement in a time interval of about two minutes taken in runs with two colliding bunches ($N_b=2$) in June 2010.  While the electromagnetic, muon, tau and jet trigger rates show a linear behavior, the MBTS\_1 rate saturates as it approaches two times the LHC revolution frequency ($N_b\times f_{LHC}\sim22$ kHz) due to pile-up. 

\clearpage

Figure~\ref{fig:beamspot} shows the \emph{beam spot} (the distribution of primary vertices in the transverse plane) reconstructed by the online beam spot algorithm using the L2 tracking and vertex finding algorithms.  Vertices are fitted from 2 or more tracks with a \pT $>$ 500 \MeV.

Figure~\ref{fig:muon} shows the L1 muon \emph{stand alone} (only muon spectrometer) trigger efficiency in the \emph{end-cap} ($|\eta|>1.05$) detector region.  The efficiency is calculated with respect to muons reconstructed offline using the muon spectrometer and inner detector tracker that have an L1 RoI within $\Delta R<0.5$.  The efficiency shows a sharp turn-on with a plateau at $\sim$90\% in the end-cap region, as expected from simulation.  

Figure~\ref{fig:tracks} shows the efficiency for EF tracking triggers as a function of the $\eta$ using offline tracks passing the tracking selection criteria for various \pT~thresholds.  The efficiency is $\sim$100\% for tracks with $\pT>1.2 \GeV$, and slightly lower for tracks with $\pT>1.0 \GeV$.   This behaviour is expected because the EF tracking triggers are configured with a minimum \pT~threshold of about 1 \GeV~due to the timing constraints.

Figure~\ref{fig:em} shows the energy spectrum of L2 trigger electromagnetic calorimeter clusters in collision events at 900~\GeV~and 7 \TeV.  The spectrum shows agreement with simulation of minimum bias events at both collision energies.

\section{Summary \& Outlook}

The ATLAS experiment has developed a flexible trigger system for efficiently collecting a diverse set of physics signals. The operation of the trigger has been very successful and the system is in the final stages of commissioning with the first LHC collision data.   No signiÞcant operational problems with the trigger have been observed.  The trigger performance is found to be consistent with expectations from simulated data.  As the instantaneous luminosity delivered by the LHC increases, the ATLAS trigger will reject a higher fraction of delivered collisions by applying increasingly stringent selection criteria and higher \pT ~thresholds.  The initial performance of the ATLAS trigger reported here gives us confidence that it will be able to handle these increasing demands in the near-term future.

\end{document}